\documentclass[12pt,prd,preprint,showpacs]{revtex4-1} 
\usepackage{graphicx}
\usepackage{amsmath}
\usepackage{amssymb}
\usepackage{epstopdf}
\usepackage{url}
\usepackage{hyperref}
\usepackage[footnotesize]{caption}
\usepackage[usenames,dvipsnames]{color}
\begin{document}

  \title{Stability of Ho\v{r}ava-Lifshitz Black Holes in the Context of AdS/CFT}
   \author{Yen Chin Ong$^{1,3}$}
     \email{d99244003@ntu.edu.tw}
   \author{Pisin Chen$^{1,2,3,4}$}
     \email{pisinchen@phys.ntu.edu.tw}
     \affiliation{%
1. Graduate Institute of Astrophysics, National Taiwan University, Taipei, Taiwan 10617\\
2. Department of Physics, National Taiwan University, Taipei, Taiwan 10617\\
3. Leung Center for Cosmology and Particle Astrophysics, National Taiwan University, Taipei, Taiwan 10617\\
4. Kavli Institute for Particle Astrophysics and Cosmology, SLAC National Accelerator Laboratory, Stanford University, Stanford, CA 94305, U.S.A.
}%

\begin{abstract}
The anti--de Sitter/conformal field theory (AdS/CFT) correspondence is a powerful tool that promises to provide new insights toward a full understanding of field theories under extreme conditions, including but not limited to quark-gluon plasma, Fermi liquid and superconductor. In many such applications, one typically models the field theory with asymptotically AdS black holes. These black holes are subjected to stringy effects that might render them unstable. Ho\v{r}ava-Lifshitz gravity, in which space and time undergo different transformations, has attracted attentions due to its power-counting renormalizability. In terms of AdS/CFT correspondence, Ho\v{r}ava-Lifshitz black holes might be useful to model holographic superconductors with Lifshitz scaling symmetry. It is thus interesting to study the stringy stability of Ho\v{r}ava-Lifshitz black holes in the context of AdS/CFT. We find that uncharged topological black holes in $\lambda=1$ Ho\v{r}ava-Lifshitz theory are nonperturbatively stable, unlike their counterparts in Einstein gravity, with the possible exceptions of negatively curved black holes with detailed balance parameter $\epsilon$ close to unity. Sufficiently charged flat black holes for  $\epsilon$ close to unity, and sufficiently charged positively curved black holes with $\epsilon$ close to zero, are also unstable.  The implication to the Ho\v{r}ava-Lifshitz holographic superconductor is discussed.
\end{abstract}


\maketitle
\flushbottom

\section{Introduction}

Anti--de Sitter/conformal field theory (AdS/CFT) correspondence  \cite{Maldacena}  \cite{Gubser1}  \cite{Witten1} has been employed to study various strongly coupled field theories, the idea being that such field theory on $(d-1)$-dimensional boundary corresponds dually to black hole physics in $d$-dimensional bulk in which the strings are weakly coupled, and so calculations in the bulk can be done semiclassically. The various applications of AdS/CFT correspondence include: quark-gluon plasma  \cite{Chamblin}  \cite{McInnes1}  \cite{McInnes2}, holographic superconductor  \cite{Hartnoll}  \cite{Debu}(for introductory reviews, see  \cite{Christopher} and  \cite{Horowitz} as well as the references therein) and holographic metals  \cite{Subir1}  \cite{Subir2}  \cite{Puletti}. See also  \cite{Hartnoll2} for a useful review. 

In the case of quark-gluon plasma, the plasma phase has nonzero minimum temperature, which means that the black hole dual to the theory should also have minimal allowed temperature bounded \emph{away} from zero due to a form of instability called the Seiberg-Witten instability  \cite{SeibergWitten}, which is closely related to the electrical charge carried by the black holes and the existence of branes in the AdS bulk. In the case of modeling superconductors with black holes, it is typical to consider not just electrical charge, but also scalar fields in the black hole spacetime. This means that the issue of stability becomes much more complicated and intricate. In view of recent applications of AdS/CFT to holographic superconductors using a Ho\v{r}ava-Lifshitz black hole by  \cite{cai-zhang}  \cite{jwc}  \cite{Setare}, we feel that the stability of Ho\v{r}ava-Lifshitz black holes in the context of AdS/CFT needs to be studied. 

We should comment at this point that Ho\v{r}ava-Lifshitz gravity is \emph{not} a string theory; it is not even a relativistic theory. Therefore one might wonder whether one can apply AdS/CFT, which is a string-inspired technique, to Ho\v{r}ava-Lifshitz gravity. This is especially a concern in our work since the Seiberg-Witten instability requires the existence of branes. We nevertheless feel that it is worth studying the consequences of applying Seiberg-Witten instability to Ho\v{r}ava-Lifshitz black holes, for several reasons. First, as mentioned above, AdS/CFT has already been applied to the study of holographic superconductors in various works  \cite{cai-zhang}  \cite{jwc}  \cite{Setare}, and seems to have yielded reasonable results. Closely related are the recent efforts to build the holographic superconductors in the bulk backgrounds with Lifshitz/Schr\"odinger scaling symmetry, according to the AdS/NCFT (nonrelativistic conformal field theory) correspondence  \cite{Zingg}  \cite{Sin}  \cite{Cremonesi}. Therefore it is not too far-fetched to consider Ho\v{r}ava-Lifshitz theory as a gravitational dual to some field theories  \cite{Tatsuma}. Second, it is possible that Ho\v{r}ava-Lifshitz gravity can be formulated in a string-theoretic way, or string theory may be modified to include Lorentz breaking  \cite{Kluson}  \cite{Kluson2}. Furthermore, as commented by Andrew Strominger \emph{et al}. in \cite{Strominger}, any consistent theory of gravity should behave a lot like string theory. In other words, extended objects like branes probably arise in \emph{any} consistent theory of quantum gravity, including nonrelativistic theories  \cite{horava0}  \cite{Kluson3}  \cite{Nabamita}  \cite{Sekhar}. In addition, AdS/CFT correspondence is also likely to occur in any quantum theory of gravity, as the existence of holographic dualities is not contingent on the validity of string theory \cite{Strominger} \cite{Strominger2}. See also \cite{Laurent} for a related discussion. To push Strominger's ideology further, we should expect that something similar (if not identical) to Seiberg-Witten instability is very likely to arise in \emph{any} theory that involves extended objects (e.g. branes) propagating in asymptotically AdS spacetimes. Therefore Seiberg-Witten instability or another qualitatively similar instability is likely to be a feature of the Ho\v{r}ava-Lifshitz version of AdS/CFT, assuming that this indeed exists, as suggested by the sensible results obtained from applying AdS/CFT techniques to studying holographic superconductor in Ho\v{r}ava-Lifshitz gravity.

Therefore, assuming that Ho\v{r}ava-Lifshitz gravity is correct, \emph{and} gravity/gauge correspondence not too different from AdS/CFT indeed exists in this theory, we hope that our work of applying Seiberg-Witten instability to a Ho\v{r}ava-Lifshitz black hole should be qualitatively, if not quantitatively, correct. In other words, while it is far from clear that Ho\v{r}ava-Lifshitz theory is compatible with string theory and therefore with AdS/CFT, we feel that the idea is worth pursuing, at least by exploring the consequences of such a premise. 

In Sec. II, we will review Seiberg-Witten stability (or the lack thereof) and its application to asymptotically AdS black holes, which can have nontrivial topology (See e.g. \cite{Birmingham} and \cite{Stefan}). In Sec. III, we will give a quick review of Ho\v{r}ava-Lifshitz topological black holes and some of their properties. In Sec. IV, we shall turn our attention to our original goal, namely the study of Seiberg-Witten stability of topological black holes in Ho\v{r}ava-Lifshitz gravity. We then conclude with some discussions in Sec. V.

\section{Instabilities in Asymptotically Anti--de Sitter Space}

In general relativity, $(n+1)$-dimensional Reissner-Nordstr\"om-AdS black holes take the form 
\begin{equation}ds^2=-f dt^2 + f^{-1}dr^2 + r^2 d\Omega_k^2\end{equation} where
\begin{equation}
f=k-\frac{16\pi M}{(n-1)\Gamma_k r^{n-2}} + \frac{8\pi Q^2}{(n-1)(n-2)\Gamma_k^2 r^{2n-4}},
\end{equation}
in which $f=f(M,Q,k,r)$ and $\Gamma_k$ denotes the area of compact space with $r=1$. Throughout this paper we will assume the charge $Q>0$ for simplicity. In $(3+1)$-dimension, we thus have, for a flat black hole,

\begin{equation}\label{f(r)}
f=\frac{r^2}{L^2} - \frac{2M}{\pi K^2 r} + \frac{Q^2}{4\pi^3 K^4 r^2}
\end{equation}
where $K$ is a continuous parameter defined by $\Gamma_0(T^2)=4\pi^2K^2$, $-\infty < K <\infty$. This also defines $K$ so that for arbitrary compact flat 2-manifolds, $K$ becomes a measure of the overall relative size of the space. 

In the study of quark-gluon plasma in flat spacetime  \cite{McInnes1}  \cite{McInnes2}, the black holes considered are those of $(4+1)$-dimension. It was shown that stringy effects can and do render charged black holes unstable as more and more electrical charges are deposited into the horizon. Furthermore, this instability occurs \textit{before the extremal limit is reached}  \cite{McInnes1}. This type of nonperturbative instability is known as \emph{Seiberg-Witten instability}, which was first pointed out in  \cite{SeibergWitten}. For an illuminating introductory account and examples of the Seiberg-Witten instability, see  \cite{Kleban}. Here we only briefly explain the idea.

Let $M$ be a smooth Riemannian manifold equipped with a metric $g$ and dimensionality $n > 2$. Suppose the scalar curvature of $M$ is $R$. Under the conformal transformation $g \to f^{\frac{n-2}{4}}g$ for some positive function $f$ defined on $M$, the Ricci scalar transforms as (corollary-1.161 of  \cite{Besse}, p. 59)
\begin{equation}
R \to g^{-\frac{n+2}{n-2}} \left(Rf + \frac{4(n-1)}{(n-2)} \Delta(f)\right)
\end{equation}
where $\Delta(f)$ is the usual Laplacian, which in local coordinates is defined by
\begin{equation}
\Delta(f) = \frac{1}{\sqrt{\text{det}(g_{kl})}} \frac{\partial}{\partial x^j}\left(\sqrt{\text{det}(g_{kl})} g^{ij} \frac{\partial f}{\partial x^i}\right).
\end{equation}
We define
\begin{equation} \displaystyle
Y[f] = \frac{\int{_M dx^n \sqrt{g} \left(Rf^2 + \frac{4(n-1)}{n-2}(\partial f)^2\right)}}{\left(\int{_M dx^n \sqrt{g} f^{\frac{2n}{n-2}}}\right)^\frac{n-2}{n}}.
\end{equation}
This quantity is defined for the conformal class of metric $[g]$. We then define the \textit{Yamabe invariant} as the infimum of all the $Y[f]$'s. Note that for $n=4$ and $f=1$, this is just a renormalized Einstein-Hilbert action.

Recall that for pure anti--de Sitter space, a bulk scalar field can have \textit{negative} squared mass, as long as it is \textit{not too negative}. To be precise, bulk scalar in $(n+1)$-dimensional AdS with mass satisfying $ m^2 \geq -n^2/4$ is allowed. Equivalently the spectrum of the Laplacian is continuous on $\left[n^2/4,\infty\right)$, having no other eigenvalue below $n^2/4$. This is the famous Breitenlohner-Freedman bound  \cite{BF}. 

If we have normalizable modes $\phi$ such that $(-\Delta + m^2)\phi = \lambda \phi$, where $\lambda < 0$ and $m$ is the mass of the scalar, then we have so-called \textit{perturbative instability}  \cite{Kleban}. This does not happen if the Breitenlohner-Freedman bound holds, e.g. in pure AdS space. However some asymptotically AdS spaces or quotients of AdS can have discrete eigenvalues below $n^2/4$. This is only possible if the \textit{Yamabe invariant} of the conformal boundary is negative  \cite{Lee}.

On the other hand, Seiberg-Witten instability is a \textit{nonperturbative} instability. The \emph{Seiberg-Witten action} is defined on the Euclidean spacetime obtained after Wick rotation by 
\begin{equation}\label{SWaction}
S=\Theta (\text{Brane Area}) - \mu (\text{Volume Enclosed by Brane})
\end{equation}
where $\Theta$ is related to the tension of the brane and $\mu$ relates to the charge enclosed by the brane due to the background antisymmetric tensor field. This brane is essentially a probe that allows us to study the background fields and geometry of the bulk. Like test particles, a probe brane is assumed not to disturb the bulk geometry and background fields. Seiberg and Witten have shown very generically that nonperturbative instability occurs when the action becomes negative due to uncontrolled brane productions. Brane-anti-brane pairs are always spontaneously created from the AdS vacuum. In analogy to the Schwinger effect in QED  \cite{Schwinger}, the rate of brane-anti-brane pair production is proportional to $\exp(-S)$, where $S$ is the Seiberg-Witten action. Thus, if $S$ is negative, the AdS vacuum nucleates brane-anti-brane pairs at an exponentially large rate instead of being exponentially suppressed. This disturbs the background geometry so much that the spacetime is no longer described by the metric that we started with, i.e. the original spacetime is not stable if such brane-antibrane production is exponentially enhanced due to the reservoir of negative action. Seiberg-Witten instability occurs precisely if the Seiberg-Witten action becomes negative at large $r$ ``close'' to the boundary, which is equivalent to the boundary having negative scalar curvature  \cite{SeibergWitten}. To understand this picture in terms of brane and antibrane dynamics in a Lorentzian picture in more details, see  \cite{Barbon}. We also remark that in analogy with the Schwinger effect, there will be a large backreaction once the brane-antibrane pairs are copiously produced, in such a way that the background geometry will evolve in response to the branes and the Seiberg-Witten instability condition might cease to hold (i.e. the action might eventually become positive). In other words, one might argue that such instability is \emph{self-limiting}. However this is \emph{not} always the case. To see why this is so, we need to understand the Seiberg-Witten instability from the dual field picture, i.e. the conformal boundary.

The conformal Laplacian of a compact manifold with metric $g$ with conformal structure is defined by
\begin{equation}
L_g \equiv -\Delta_g + \frac{n-2}{4(n-1)}R(g)
\end{equation}
where $\Delta_g$ is the usual Laplacian where we have emphasized its dependence on the metric $g$. The conformal Laplacian is an elliptic operator with a discrete real spectrum bounded from below. Suppose $\lambda_1$ is its first eigenvalue, then the field theory defined on the boundary is stable if $\lambda_1 > 0$ and unstable if $\lambda_1 < 0$. The case for $\lambda_1 = 0$ is more delicate and requires more analysis. Note that the eigenvalues $\lambda_i = \lambda_i (g)$ are also dependent on the metric.

A classical problem in differential geometry called \textit{Yamabe problem} is the following: 
\begin{quote}
Given a smooth, compact manifold $M$ of dimension $n > 2$ with a Riemannian metric $g$, does there exist a metric $\tilde{g}$ conformal to $g$ such that the scalar curvature of $\tilde{g}$ is constant?
\end{quote}
The answer is affirmative as shown by Schoen  \cite{Schoen}. Therefore, there indeed exists such $\tilde{g}$ so that the scalar curvature $R(\tilde{g})$ is constant. It is in fact equal to
\begin{equation}
R(\tilde{g}) = \frac{4(n-1)}{n-2}\lambda_1(\tilde{g})
\end{equation}
so that
\begin{equation}
L_{\tilde{g}} = -\Delta_{\tilde{g}} + \lambda_1(\tilde{g}).
\end{equation}
Therefore stability depends on the sign of the scalar curvature at the boundary. Indeed, in the case where Seiberg-Witten instability occurs, the boundary has negative scalar curvature, and thus correspondingly the field theory is unstable due to a negative squared mass scalar field in the dual field description.  Note that pure anti--de Sitter space has positively curved conformal infinity and so is stable in the Seiberg-Witten sense. We note from Eq.(\ref{SWaction}) that area contributes positively to the action, but the volume contributes negatively. This means that the volume enclosed by the brane must not grow too rapidly relative to the area, otherwise instability will eventually occur. Assuming supersymmetry in the bulk (this is not entirely impossible even for Ho\v{r}ava-Lifshitz gravity  \cite{Xue}), the amount of charge is bounded above. Clearly the most dangerous case is when $\mu$ is maximum; this is the BPS case in which in $(n+1)$--dimension is given by $\mu_{\text{BPS}}=n\Theta/L$. Indeed this analysis is not strictly restricted to the AdS bulk, but also holds for an arbitrary Einstein manifold of negative curvature and conformal boundary  \cite{SeibergWitten}. 

We remark that both perturbative and nonperturbative (Seiberg-Witten) instabilities do not occur if the Yamabe invariant is positive, or equivalently, if the conformal boundary has positive scalar curvature. However, see the subsequent discussion for the case where the Seiberg-Witten action is negative for some finite range yet still asymptotically divergent. 

Now coming back to a previous remark that Seiberg-Witten instability is \emph{not} always self-limiting. This is the consequence of the fact that  \emph{there exist
compact manifolds on which it is impossible to define a Riemannian metric of positive or zero scalar curvature}  \cite{brett0}. For such cases, the AdS bulk is unstable due to emission of brane-antibrane pairs and will \emph{remain unstable no matter how the metric is distorted due to backreaction}. This is the case for black holes with negatively curved horizon in general relativity: once brane-antibrane pairs are produced, nothing can stop the instability; no matter how the branes deform the spacetime, the scalar curvature at infinity can never become everywhere positive or zero  \cite{Brett00}. This is also clear from the fact that the Seiberg-Witten action is \emph{unbounded below} in this case, as we will discuss in more detail in Sec. IV. Therefore, whether instability is self-limiting or not depends on the \emph{topology} of the underlying manifold.

We should at this point stress that the stability issue discussed depends on whether the theory is classical or semiclassical. For example \emph{classically} Reissner-Nordstr\"om-AdS black holes are gravitationally stable  \cite{Roman} despite the existence of a thermodynamically unstable parameter range; but Gubser-Mitra instability  \cite{GubserMitra} occurs for Reissner-Nordstr\"om-AdS black holes in $\mathcal{N}=8$ gauged supergravity, due to tachyon mode of the scalar field which coupled to the system which causes thermodynamically unstable black holes to be also dynamically unstable. The Seiberg-Witten instability is likewise \textit{not} a classical effect. 

In the case of $k=+1$ and $k=-1$ black holes, the Seiberg-Witten stability issue is straightforward: positively curved black holes are stable while negatively curved black holes are always unstable in the Seiberg-Witten sense  \cite{McInnes3}. Flat black holes, however, are marginally stable: the Seiberg-Witten action asymptotes to a value linearly proportional to its mass, and adding charge lowers the action such that it eventually becomes negative at large $r$, rendering the black hole unstable. To be precise, the instability occurs as the amount of charge reaches  about $95.8\%$ of the extremal charge for charged (4+1)-dimensional black holes in anti--de Sitter space \cite{McInnes1}. Therefore, the Seiberg-Witten instability provides an explanation of why quark-gluon plasma cannot be arbitrarily cold -- namely the dual black hole cannot be too cold if it is to be stable. 

However, as we have pointed out, applications of AdS/CFT correspondence are becoming wider by the day, and in some of these cases, the field theory is not bounded away from zero. One such example is the Fermi liquid. The black hole dual to the Fermi liquid might be a black hole with dilaton charge or a relative to it  \cite{GubserRocha}. Indeed dilaton black holes have been extensively studied for their holography and applications in AdS/CFT correspondence  \cite{Goldstein}  \cite{Chiangmei}  \cite{Cadoni}  \cite{Meyer}. In the case where the dual field theory can be arbitrarily cold or even reach zero temperature, one wishes that the corresponding black holes were stable in the Seiberg-Witten sense. The dilaton hair is not a fundamental ``hair'' since it couples to the Maxwell field. For a flat dilaton black hole, at least for those with coupling strength $\alpha=1$, the Seiberg-Witten action remains positive as the electrical charge increases  \cite{yenchin}. Indeed, for typical fixed charge $Q_1$, increasing the charge to $Q_2 > Q_1$ makes the action starts out with $S(Q_2) < S(Q_1)$ initially, but subsequently takes over at some finite value of $r=R$ so that $S(Q_2) > S(Q_1)$ for all $r \geq R$. The value of $r$ in which this takeover occurs decreases with increasing charge.

On a similar note, some types of Ho\v{r}ava-Lifshitz black holes have been studied as being dual to superconductors with Lifshitz scaling symmetry  \cite{cai-zhang}  \cite{jwc}  \cite{Setare}, and we would like to explore the stability of these black holes when we take into account the influence of branes to the AdS bulk geometry semiclassically. However before we discuss this issue, let us first review the properties of topological black holes in Ho\v{r}ava-Lifshitz gravity.

\section{Topological Black Holes in Ho\v{r}ava-Lifshitz Gravity}

The idea of Ho\v{r}ava-Lifshitz gravity originated from the study of the longstanding problem regarding nonrenormalization of general relativity. It is suggested that nonrenormalizability implies that general relativity is only an effective theory which will break down in the high-energy regime. As an effective theory then, the curvature scalar in the Einstein-Hilbert action is probably only the first of many higher order curvature terms. An attempt to renormalize gravity by naively introducing higher order terms however is problematic because these terms have derivatives of both spatial and temporal kinds, and we know that from a field theoretical point of view, higher order time derivatives lead to problems like ghost degrees of freedom which render the theory nonunitary. Therefore, the idea of Ho\v{r}ava-Lifshitz gravity  \cite{Horava1} is to break Lorentz invariance so that we can have higher spatial derivative terms yet no higher time derivative terms. In other words time and space are \emph{not} to be treated on equal footing. This makes the theory power-counting renormalizable if space and time transform as $\textbf{x} \to b\textbf{x}, t \to b^3t$ for some constant $b$. Note that in this construction, Lorentz invariance can be recovered in the infrared limit where $\lambda \to 1$, so that the theory will reduce to well-tested general relativity. For a timely review, see  \cite{Thomas}.

In the well known (3+1)-dimensional Arnowitt-Deser-Minser (ADM) formalism  \cite{ADM}, the spacetime metric can be written as 
\begin{equation}
ds^2 = -N^2 dt^2 + g_{ij}(dx^i + N^i dt)(dx^j + N^j dt),
\end{equation}
where $N$ is the lapse function and $N^i$ is the shift vector.

The Einstein-Hilbert (EH) action is
\begin{equation}
S_{\text{EH}}=\frac{1}{16\pi G} \int d^4x \sqrt{g}N\left[K_{ij}K^{ij} - K^2 + R -2\Lambda \right],
\end{equation}
where $K_{ij}$ is the extrinsic curvature 
\begin{equation}
K_{ij} = \frac{1}{2N}\left(\dot{g}_{ij}-\nabla_i N_j - \nabla_j N_i \right),
\end{equation}
with the dot denoting derivative with respect to time. Note that the covariant derivative is a spatial one.

The action of the Ho\v{r}ava-Lifshitz gravity is, with $g$ denoting the determinant of the spatial metric $g_{ij}$,

\begin{equation}
I = \int dt d^3x \sqrt{g} [\mathcal{L}_K + \mathcal{L}_V]
\end{equation}
where
\begin{equation}
\mathcal{L}_K = \dfrac{2}{\kappa^2}(K_{ij}K^{ij} - \lambda K^2) 
\end{equation}
is the kinetic term and

\begin{equation}
\mathcal{L}_V = \dfrac{\kappa^2 \mu^2 (\Lambda R - 3\Lambda^2)}{8(1-3\lambda)} 
+ \dfrac{\kappa^2 \mu^2 (1-4\lambda)}{32(1-3\lambda)}R^2 - \dfrac{\kappa^2\mu^2}{8}R_{ij}R^{ij} 
+ \dfrac{\kappa^2 \mu}{2 \omega^2}\epsilon^{ijk}R_{il}\nabla_j R^l_k - \dfrac{\kappa^2}{2\omega^4}C_{ij}C^{ij}
\end{equation}
is the potential term determined by what is known as a \emph{detailed balance condition}, which is inspired from condensed matter physics. 

Here $\kappa^2, \lambda, \omega, \mu$ and $\Lambda$ are all parameters of the theory, while 
\begin{equation}
C^{ij}=\epsilon^{ikl}\nabla_k \left(R^j_l - \frac{1}{4} R \delta_l^j\right)
\end{equation}
is the Cotton tensor. Of particular importance is the running coupling $\lambda > 1/3$, which at IR limit is expected to flow to $\lambda=1$ where general relativity is recovered.

We remark that the detailed balance condition is not an essential feature of the theory, but it drastically reduces the number of terms one needs to consider. The Lagrangian in the theory without detailed balance condition, with $0 < \epsilon \leq 1$, takes the following form:
\begin{equation}
\mathcal{L} = \mathcal{L}_0 + (1-\epsilon^2)\mathcal{L}_1
\end{equation}
where
\begin{equation}
\mathcal{L}_0 = \sqrt{g} N \left[\dfrac{2}{\kappa^2}(K_{ij}K^{ij} - \lambda K^2) + \dfrac{\kappa^2 \mu^2 (\Lambda R - 3\Lambda^2)}{8(1-3\lambda)} \right]
\end{equation}
and 

\begin{equation}
\mathcal{L}_1=  \sqrt{g} N \left[ \dfrac{\kappa^2 \mu^2 (1-4\lambda)}{32(1-3\lambda)}R^2 - \dfrac{\kappa^2\mu^2}{8}R_{ij}R^{ij} 
+ \dfrac{\kappa^2 \mu}{2 \omega^2}\epsilon^{ijk}R_{il}\nabla_j R^l_k - \dfrac{\kappa^2}{2\omega^4}C_{ij}C^{ij}\right].
\end{equation}

We call $\epsilon$ the \emph{detailed balance parameter}. Detailed balance condition is obtained when $\epsilon=0$, while general relativity is recovered when $\epsilon=1$. One should note that however in application to our physical Universe, the detailed balance condition is tightly constrained from observations and is actually disfavored, although not completely ruled out  \cite{Chien-I}  \cite{Horatiu}. 

The speed of light in Ho\v{r}ava-Lifshitz theory is not fundamental but is given by:
\begin{equation}
c=\frac{\kappa^2 \mu}{4} \sqrt{\frac{\Lambda}{1-3\lambda}},
\end{equation}
while Newton's constant is given by 
\begin{equation}
G = \frac{\kappa^2 c^2}{16\pi(3\lambda -1)}.
\end{equation}
Here $\Lambda$ is related to the cosmological constant (CC) $\Lambda_{\text{CC}}$ by $\Lambda = 2\Lambda_{\text{CC}}/3 $. We note from the square root in the expression for $c$ that the cosmological constant must thus be \textit{negative}, although it is argued that it can be made positive via analytic continuation  \cite{LMP}. 

Spherically symmetric black hole solutions in Ho\v{r}ava-Lifshitz theory have been found  \cite{LMP}, followed shortly by solutions of topological black holes  \cite{CCO}. See also  \cite{Ehsan} and  \cite{Konoplya}. We shall see that, as has been pointed out in these works, even in the limit $\lambda=1$, these black holes have very different behaviors than their counterparts in Einstein's theory of general relativity (see also  \cite{Myung-Kim}). We will focus on the case $\lambda=1$ in the following discussion. However before proceeding, we would like to remark that there is much we do not yet understand about black hole solutions in Ho\v{r}ava-Lifshitz theory. For example, the horizon radii generically depend on the energies of test particles  \cite{Greenwald}, that is, it is not clear ``when and for whom they are black''  \cite{Kiritsis}. Furthermore, many black hole--like solutions might not be black holes in the usual sense due to different dispersion relations in the Ho\v{r}ava-Lifshitz theory  \cite{Greenwald}. Nevertheless, since naive application to holographic superconductors seems to yield reasonable results, we shall assume the validity of this approach and proceed to study its consequences.

For comparison, let us first look at charged topological AdS black holes in Einstein's general relativity (GR): with metric $ds^2 = -f dt^2 + f^{-1}dr^2 + r^2 d\Omega^2$ where in $(3+1)$-dimension, 
\begin{equation}
f(M,Q,r)=k - \frac{8\pi M}{r\Gamma_k} + \frac{4\pi Q^2}{\Gamma_k^2 r^2} + \frac{r^2}{L^2}.
\end{equation}
The cosmological constant is $\Lambda_{\text{CC}}=-3/L^2$, where $L$ is the length scale of the anti--de Sitter space. Thus $\Lambda = -2/L^2$. 

In the absence of charge, 
\begin{equation}
f(Q=0,M,r)=k-\frac{8\pi M}{r\Gamma_k} + \frac{r^2}{L^2},
\end{equation}
and so the temperature 
\begin{equation}
T=\frac{f'(r_+)}{4\pi}
\end{equation}
where $r_+$ denotes the horizon, is given by
\begin{equation}\label{temperature-Einstein}
T_{\text{GR}}=\frac{\sqrt{-\Lambda}}{8\pi x_+}(3x_+^2 + 2k)
\end{equation}
where we have defined $x \equiv r\sqrt{-\Lambda}$.

For Ho\v{r}ava-Lifshitz topological black holes with $\lambda=1$, we have  \cite{CCO}

\begin{equation}
f(x) = k + \frac{x^2}{1-\epsilon^2} - \frac{\sqrt{\epsilon^2 x^4 + (1-\epsilon^2)c_0 x}}{1-\epsilon^2}
\end{equation}
where
\begin{equation}
c_0 = \frac{x_+^4 + 2kx_+^2 + (1-\epsilon^2)k^2}{x_+}.
\end{equation}
Expanding the square root terms as a power series, we can see that

\begin{equation}
f(x) = k + \frac{x^2}{1+\epsilon} - \frac{c_0}{2\epsilon x} + O\left(\frac{1-\epsilon^2}{x^4}\right).
\end{equation}
Thus, we see that for $\epsilon \to 1$, the higher order terms vanish and we end up with
\begin{equation}
f(x) = k + \frac{x^2}{2} - \frac{c_0}{2x}
\end{equation}
which recovers the topological uncharged black hole solutions in AdS. Furthermore even for $\epsilon \neq 1$, for large $x$ (i.e. large $r$), the solution is again asymptotically AdS.

In the presence of electrical charge, we have  \cite{CCO}
\begin{equation}\label{fwc}
f(x) = k + \frac{x^2}{1-\epsilon^2} - \frac{\sqrt{\epsilon^2 x^4 + (1-\epsilon^2)\left(c_0 x - \frac{q^2}{2}\right)}}{1-\epsilon^2}
\end{equation}
where the charge parameter $q$ is related to the previous charge $Q$ by
\begin{equation}
Q=\frac{\kappa^2\mu^2\Gamma_k\sqrt{-\Lambda}}{16}q
\end{equation}
and
\begin{equation}
M=\frac{\kappa^2 \mu^2 \Gamma_k \sqrt{-\Lambda}}{16}c_0.
\end{equation}
Here
\begin{equation}\label{c0}
c_0 = \frac{x_+^4 + 2kx_+^2 + (1-\epsilon^2)k^2 + \frac{q^2}{2}}{x_+}.
\end{equation}

In the $\epsilon \to 1$ limit, we recover the usual AdS charged topological black hole solutions
\begin{equation} \label{f(x) from HL}
f(x) = k + \frac{x^2}{2} - \frac{c_0}{2x} + \frac{q^2}{4x^2}.
\end{equation}
The Hawking temperature is  \cite{jwc}
\begin{equation}
T = \frac{\sqrt{-\Lambda}\left[3x_+^4 + 2kx_+^2 - (1-\epsilon^2)k^2 - \frac{q^2}{2}\right]}{8\pi x_+ [x_+^2 + (1-\epsilon^2)k]}.
\end{equation}

For convenience of discussion, we give explicit forms of the relevant equations for the detailed balance case:

As $\epsilon=0$, chargeless black holes in Ho\v{r}ava-Lifshitz gravity are given by  \cite{CCO}:
\begin{equation} \label{detailed-balanced-f}
f(r)=k+x^2-\sqrt{c_0 x}, 
\end{equation}
where 
\begin{equation}
c_0= \frac{k^2 + x_+^4 + 2kx_+^2}{x_+} > 0.
\end{equation}
The horizon has constant curvature $2k$, where $k=+1,0,-1$. In this case, it was shown that the temperature and entropy take the form  \cite{CCO}
\begin{equation} \label{temperature-neutral-detailed-balanced}
T = \frac{\sqrt{-\Lambda}}{8 \pi x_+}(3x_+^2 - k), 
\end{equation}
and
\begin{equation}
S= \frac{c^3}{4G}\left(A - \frac{k\Gamma_k}{\Lambda} \ln \frac{A}{A_0}\right),
\end{equation}
respectively, where $A_0$ is a constant of dimension length squared, which cannot be determined without knowing some details of quantum gravity. Thus we notice that the thermodynamics of black holes in Ho\v{r}ava-Lifshitz theory is very different from that in general relativity: the entropy is no longer a quarter of the horizon area, but contains a correction term which scales as a logarithm of the area. This is not very surprising since in modified gravity, in general, the entropy is not exactly equal to a quarter of horizon area (see, e.g. lesson-6 of  \cite{Padmanabhan}). 

An interesting feature to note, as pointed out in  \cite{CCO} and  \cite{CCS}, is that the behavior of the temperature of black holes in Ho\v{r}ava-Lifshitz theory is opposite to that of the black holes in general relativity, i.e. that $k=+1$ black holes in one theory behave like $k=-1$ black holes in the other theory and vice versa. This can be seen from the opposite sign in front of the $k$ term in the temperature expressions of the two theories [Eqs.(\ref{temperature-Einstein}) and  Eq.(\ref{temperature-neutral-detailed-balanced})]. The ``duality'' is not exact. For example, we know that there exists minimum allowed temperature for $k=1$ black holes in general relativity. This is given by the turning point of Eq.(\ref{temperature-Einstein}) with $k=1$, which occurs at $x_+ = \sqrt{2/3}$, or equivalently at $r=L/{\sqrt{3}}$. Thus black holes under a certain critical temperature are unstable. Indeed, there exists a Hawking-Page transition for $k=1$ black holes \cite{Hawking}. This is however not the case for $k=-1$ Ho\v{r}ava-Lifshitz black holes  \cite{CCO}. This is because with $k=-1$, Eq. (\ref{detailed-balanced-f}) evaluated on the horizon gives 
\begin{equation}
x_+^2 - \sqrt{c_0 x_+} = 1
\end{equation}
which enforces $x_+ \geq 1$. But the turning point of Eq.(\ref{temperature-neutral-detailed-balanced}) occurs at $x_+ = \sqrt{1/3}$. Since $x_+|_{\text{min}} > \sqrt{1/3}$, the unstable phase for $k=-1$ Ho\v{r}ava-Lifshitz black holes does not arise and so these black holes are thermodynamically stable. In fact, all uncharged topological black holes are thermodynamically stable in Ho\v{r}ava-Lifshitz theory. This result is for $\lambda=1$ only. Things are somewhat different for general $\lambda$  \cite{CCO2}. We remind the reader that for $k=0$ black holes in general relativity, there is no Hawking-Page phase transition into an AdS background, although there can be transition into Horowitz-Myers soliton  \cite{HM} \cite{Surya}, which we will not discuss here.

In the presence of electrical charge, we have, by setting $\epsilon=0$ to in Eq.(\ref{fwc}),
\begin{equation}
f(r) = k + x^2 - \sqrt{c_0 x - \frac{q^2}{2}}
\end{equation}
where
\begin{equation}
c_0 = \frac{2k^2 + q^2 + 4k x_+^2 + 2x_+^4}{2x_+} > 0.
\end{equation}
The temperature is
\begin{equation}
T = \frac{6x_+^4 + 4kx_+^2 - 2k^2 -q^2}{16 L^2 \pi x_+(k + x_+^2) }.
\end{equation}
Thus we see that the extremal limit $T=0$ is achieved at 
\begin{equation}
x_{E} = \sqrt{-\frac{k}{3} + \frac{\sqrt{8k^2+3q^2}}{3\sqrt{2}}}.
\end{equation}
The charge parameter thus satisfies
\begin{equation}
q^2 \leq 2(-k^2 + 2kx_+^2 + 3x_+^4)
\end{equation}
with the bound saturated for extremal black hole.

The temperature for $k=-1$ charged black holes then exhibit different behaviors from the uncharged case: there is now minimum temperature for small black holes, i.e. black holes with $0< x_+ < 1$. This minimum point occurs at $x=1/\sqrt{3}$ of the value of the charge, as long as it is nonzero. Thus charged black holes are thermodynamically stable for the $k=0$ and $k=1$ cases, and also stable for the $k=-1$ case if the black hole is sufficiently large. See  \cite{CCS} for detailed discussions.

Therefore even if the Ho\v{r}ava-Lifshitz theory can recover Einstein's theory in the IR, the \emph{solutions} might not. This is of course not the only problem with  Ho\v{r}ava-Lifshitz gravity  \cite{limiao}. Regardless of the validity of Ho\v{r}ava-Lifshitz theory as renormalizable theory of quantum gravity to describe our physical world, it is hoped that the 
black hole solutions in Ho\v{r}ava-Lifshitz theory might nevertheless be useful in application to understand superconductor-type phenomena via AdS/CFT, which we will discuss in the following sections.

\section{Holography of Ho\v{r}ava-Lifshitz Black Holes}

Ho\v{r}ava-Lifshitz black holes have been studied for their possible applications in AdS/CF; see e.g.  \cite{cai-zhang},  \cite{jwc},  \cite{Setare}. In  \cite{cai-zhang}, the authors studied the phase transition of flat Ho\v{r}ava-Lifshitz black holes by introducing a Maxwell field and a complex scalar field and found that the results are rather similar to those in the case of black holes in general relativity. They thus concluded that the superconductivity phenomenon is rather robust, insensitive to gravitational theories at hand, but rather associated with asymptotic AdS black holes. In  \cite{jwc}, the work is extended to the case without detailed balance condition.

In this section, we only consider (3+1)-dimensional charged Ho\v{r}ava-Lifshitz black holes. We do not include effects of scalar fields. However, see Sec. V for further discussion.

Following  \cite{McInnes1} and  \cite{McInnes3}, we consider the Wick-rotated version of the black hole metric with a BPS brane of tension $\Theta$ wrapping one of the $r=\text{const.}$ sections of the resulting space of Euclidean signature. The Seiberg-Witten action is then
\begin{equation}
S(r)=\Theta A - \mu V 
\end{equation}
where 
\begin{equation}
A(r) = \sqrt{g_{\tau\tau}} \int d\tau  \int r^2 d\Omega_k
\end{equation}
and 
\begin{equation}
V(r) = \int_{r_+}^r \sqrt{g_{\tau\tau}} \sqrt{g_{r'r'}} r'^2 dr' \int d\Omega_k \int d\tau
\end{equation}
are the brane area and volume enclosed by the brane, respectively. Since the brane is BPS, we have $\mu = 3\Theta/L$ in (3+1)-dimension. 

Recall that in performing Wick-rotation, the time coordinate $t$ now parametrizes a circle. We can think of $t/L$ as an angular coordinate on this circle with periodicity $P$ chosen so that the metric is not singular at $r_+$. 

The Seiberg-Witten action is

\begin{equation}\label{S for HLC}
S(x,k,q,\epsilon) = PL \Theta \Gamma_k \frac{L^2}{2}\left\{ x^2 \left[k + \frac{x^2 - \sqrt{\epsilon^2 x^4 + (1-\epsilon^2)(c_0 x-\frac{q^2}{2})}}{1-\epsilon^2}\right]^{\frac{1}{2}} - \frac{1}{\sqrt{2}} (x^3 - x_+^3) \right\}
\end{equation}
where we recall that

\begin{equation}
c_0 = \frac{k^2(1-\epsilon^2) + x_+^4 + 2kx_+^2 + q^2/2}{x_+}.
\end{equation}

For a consistency check, as mentioned in Section II, for the case $\epsilon = 1$, we should recover Reissner-Nordsrt\"om-AdS solution of general relativity. For flat case, we then have, henceforth ignoring the overall positive multiple $PL \Theta \Gamma_k L^2/2$, 
\begin{equation}
S \propto x^2\sqrt{\frac{x^2}{2}-\frac{x\left(x_+^3+\frac{q^2}{2x_+}\right)-\frac{q^2}{2}}{2x^2}}-\frac{x^3-x_+^3}{\sqrt{2}}.
\end{equation}

With $x_+=1$, we increase the electrical charge and observe that eventually the action approaches zero at infinity for some near-extremal charge parameter $q_{\text{NE}}$. We claim that $q_{\text{NE}}^2 = 2$. Indeed, we can check that with this value, the action becomes, 
\begin{flalign}\label{S for RN}
S &\propto  x^2\sqrt{\frac{x^2}{2}-\frac{2x-1}{2x^2}} -\frac{x^3-1}{\sqrt{2}} \\
&= \frac{x}{\sqrt{2}}\sqrt{x^4-2x+1} - \frac{x^3-1}{\sqrt{2}}, 
\end{flalign}
which indeed tends to 0 as $x \to \infty$.

As shown in Fig.1, for any amount of charge exceeding $q=\sqrt{2}$, the action becomes negative at some finite $x$ and \textit{stays negative} afterward. This signals instability by brane production, and so the black holes become unstable before extremality is reached. Correspondingly then, the temperatures of such stable black holes are bounded \emph{away} from zero.

Comparing Eq.(\ref{f(r)}) and Eq.(\ref{f(x) from HL}), we see that
\begin{equation}
\frac{q^2}{4x^2} = \frac{Q^2}{4\pi^3 K^4 r^2}.
\end{equation}
Recall that $x=\sqrt{-\Lambda} r = \sqrt{2/L^2} r$; this reduces to
\begin{equation}
q^2 = -\frac{\Lambda Q^2}{\pi^3 K^4}.
\end{equation}
Therefore our result that $q_{\text{NE}}^2 = 2$ agrees with the result in  \cite{McInnes4}, from which we know that for flat black holes \emph{in general relativity} with $n$-dimensional horizon, the near-extremal charge satisfies 
\begin{equation}
\pi Q_{\text{NE}}^2 L^2 = 2^{\frac{5n-3}{n+1}}(n-1)n^{\frac{1-n}{n+1}}\left[\pi^2MKL^2\right]^{\frac{2n}{n+1}} 
\end{equation}
and the horizon with this amount of charge satisfies
\begin{equation}
r_{\text{NE}} = \left(\frac{ML^2}{2^{n-3}n\pi^{n-1}K^n}\right)^{\frac{1}{n+1}}.
\end{equation}
Our case corresponds to $n=2$, and we have set $x_+ = 1$. By setting $L=1$ and $K=1$ and varying $M$, we have indeed 
\begin{equation}
\frac{1}{\sqrt{2}}=r_{\text{NE}}= \left[\frac{ML^2}{\pi K^2}\right]^\frac{1}{3} \Rightarrow M = \frac{\pi}{\sqrt{8}}
\end{equation}
and thus
\begin{equation}
Q_{\text{NE}}^2 = 4\pi^3\left(\frac{1}{\sqrt{8}}\right)^{\frac{4}{3}} \Rightarrow q_{\text{NE}}^2 = 2
\end{equation}
as we have found previously. Therefore, we have a consistency check that we do recover the result of Seiberg-Witten instability for charged flat black holes in general relativity when $\epsilon=1$.

\begin{figure}
\begin{center}
\includegraphics[width=3.0 in]{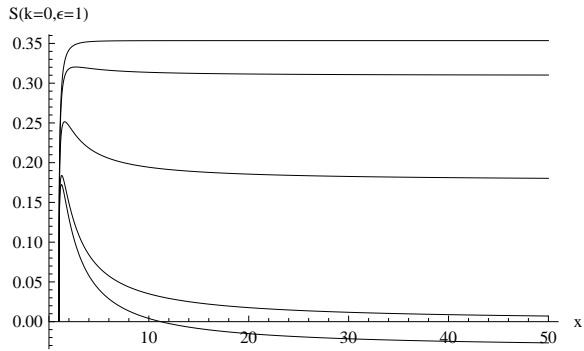}
\caption{Flat black holes with $\epsilon=1$. From top to bottom, the charge parameter values are $q=0,0.5,1,\sqrt{2},1.48$, respectively. The value $q=\sqrt{2}$ corresponds to the amount of charge where the Seiberg-Witten action becomes zero at infinity.}
\end{center}
\end{figure}

The behavior of the Seiberg-Witten action hugely depends on the value of $\epsilon$. Even for a fixed value, say $\epsilon=0$ for the detailed balance case, we see that for the uncharged case (top-left diagram in Fig.2), the action for $k=-1$ black holes is positive and even \emph{greater} than the flat and positively curved cases, which is a completely different behavior than its general relativistic counterpart in which the action for negatively curved black holes always turns to become negative at large $x$. Increasing electrical charge does lower the action of such a detailed balance
 black hole, but the asymptotic behavior does not change -- the action still tends to positive infinity at large $x$ regardless of the horizon curvature.

Note, however, that for some values of parameters, the action does become negative for some intermediate range of $x$, although it is positive for large $x$. Note that this also happens for charged $k=1$ black holes at sufficiently small $\epsilon$, although their action is always positive in the case of general relativity. The fact that the action becomes negative at some finite range does signal nonperturbative instability under brane nucleation as before, but of a milder form. One way to interpret this is as follows: brane-antibrane pairs are created at an exponential rate from the reservoir of energy where the action is negative. For the action which is negative between some finite range, brane-antibrane pairs are produced with the expense of the electrical charge $\mu$ on the brane so that eventually the action becomes \textit{less negative}. In other words the black hole spacetime which is unstable in this sense will eventually settle down below the threshold value that triggered the instability. However, when everything has settled down to a stable configuration, it is \textit{no longer the original spacetime}. It has become a ``nearby solution'' in the sense of Maldacena and Maoz  \cite{MaldacenaMaoz}. We contrast this to the case of $k=-1$ black holes and near-extremal $k=0$ black holes in general relativity in which once the action becomes negative it stays negative. In such a scenario, the spacetime is genuinely unstable because the action cannot become positive by nucleating a finite number of brane-antibrane pairs. 

Nearby solutions -- in the sense of Maldacena and Maoz may correspond to deforming the black holes in some ways, which should not be confused with so-called ``deformed'' Ho\v{r}ava-Lifshitz black holes (Kehagias-Sfetsos black holes)  \cite{KehagiasSfetsos}  \cite{CCS}. The latter refers to the situation where there exists a term in the IR modified action of Ho\v{r}ava-Lifshitz gravity which allows one to obtain a Minkowski vacuum instead of an AdS vacuum. 

Furthermore, we observe in Fig.3 that even for uncharged black holes $q=0$, the actions for black holes of different horizon curvature cross over as $\epsilon$ varies from $0$ to $1$. With $x_+$ set to 1, we observe that under detailed balance condition ($\epsilon=0$), the action for the $k=-1$ black hole is the highest, followed by $k=0$ and finally the $k=+1$ case. But eventually it becomes the other way around at $\epsilon \to 1$.

We observe that as the electrical charge increases, regardless of values of $\epsilon$ and $k$, the Seiberg-Witten action for Ho\v{r}ava-Lifshitz black holes decreases. This is the same behavior as the case in general relativity. 

We note that the actions for flat black holes remain infinite as $x \to \infty$ if $\epsilon \neq 1$; indeed from Eq.(\ref{S for HLC}), we see that, at large $x$,
\begin{equation}
S \propto x^2\left[\frac{x^2-\epsilon x^2}{1-\epsilon^2}\right]^{\frac{1}{2}}-\frac{x^3}{\sqrt{2}} = x^3\left[\frac{1}{\sqrt{1+\epsilon}}-\frac{1}{\sqrt{2}}\right].
\end{equation}
Thus for $0 \leq \epsilon < 1$, the Seiberg-Witten action always blows up at infinity \textit{independent of the charge}. 

For the $\epsilon=1$ case which corresponds to general relativity, the result above  does not hold because we have to deal with indeterminate form. Indeed the action in this case reduces to that of general relativity, that is, it asymptotes to a constant, which does \emph{depend} on the value of the charge as we have seen. Therefore there is a discontinuity in the behavior of the Seiberg-Witten action as $\epsilon$ varies from $\epsilon=0$ toward $\epsilon=1$.

For negatively curved black holes, the action at large $x$ is
\begin{equation}
S \propto x^2 \left[-1 + \frac{x^2}{1+\epsilon}\right]^{\frac{1}{2}} - \frac{x^3}{\sqrt{2}},
\end{equation}
which can be negative if and only if 
\begin{equation}
-1 + \frac{x^2}{1+\epsilon} < \frac{x^2}{2},
\end{equation}
i.e. if and only if $\epsilon > 1$ as $x \to \infty$.

\begin{figure}
\begin{center}
\includegraphics[width=5.5 in]{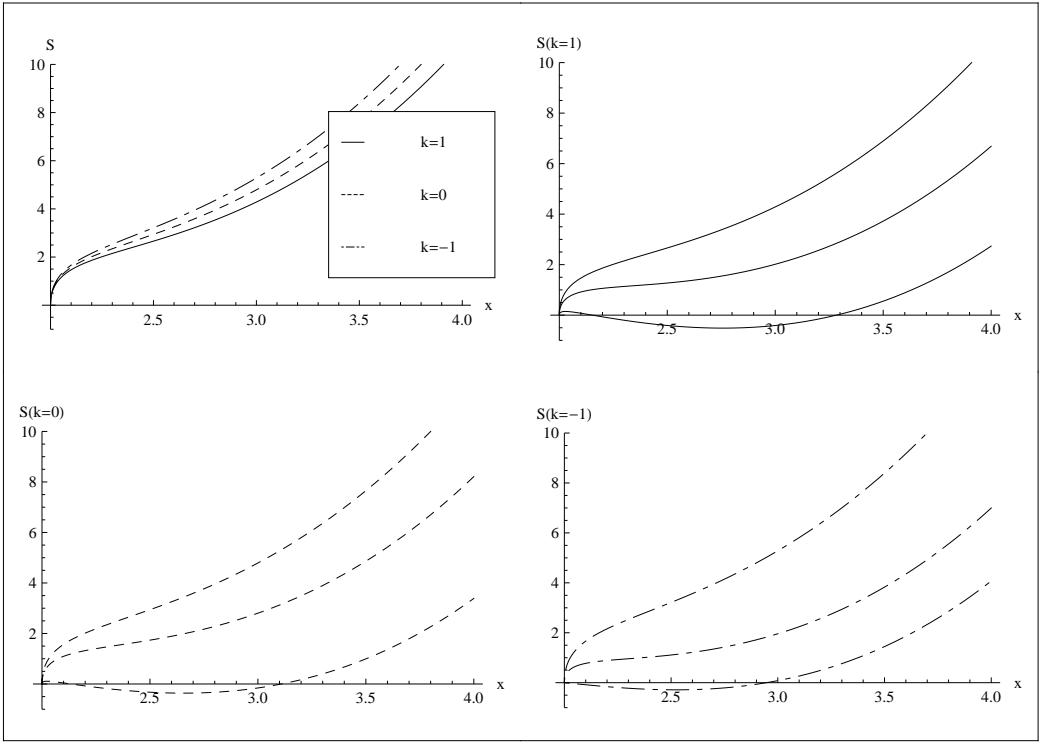}
\caption{Top left: Seiberg-Witten actions for uncharged black holes with $\epsilon=0$ where the curves from top to bottom are that of $k=-1$, $k=0$, $k=+1$, respectively. Top right, bottom left and bottom right show the behavior for the action as charge is increased (higher curves correspond to lower charge), for $k=+1,0,-1$ cases, respectively. The horizons (origins) are set to $x=2$. Note that the action for sufficiently charged $k=1$ black holes can be negative in some range of $x$.}
\end{center}
\end{figure}

Therefore similar phenomenon happens for negatively curved black holes: While the Seiberg-Witten action eventually turns over and becomes negative \emph{and stay negative} for $\epsilon=1$, the action remains positive at infinity for all values of $\epsilon < 1$. We plot the Seiberg-Witten actions of topological black holes as the charge increases in Fig.4.

In fact, we see from Fig.6 that by holding the charge fixed and increasing $\epsilon$ toward 1, the turning point of the action for $k=-1$ black holes shifts progressively toward the right, and with $\epsilon \to 1$, this turning point gets pushed all the way to infinity, i.e. it does not turn over for the case $\epsilon=1$, which recovers the case for general relativity with infinite reservoir of negative action. A similar phenomenon happens for the case of $k=0$ black holes. The implication of this observation is that the ``amount of negative reservoir'' is \emph{not bounded}. As the brane-antibranes are produced in huge numbers in the bulk, the black hole spacetime will be disturbed so much that one may reasonably worry that it might develop into true instability. That is to say, although the Seiberg-Witten at infinity is positive which usually means the field theory on the boundary is stable, this ``stability'' in the case of  Ho\v{r}ava-Lifshitz black holes should not be trusted until we have more details on how the spacetime deforms and settles down after it gets rid of the negative reservoir of action via the huge number of brane-antibrane emissions.

\begin{figure}
\begin{center}
\includegraphics[width=5.7 in]{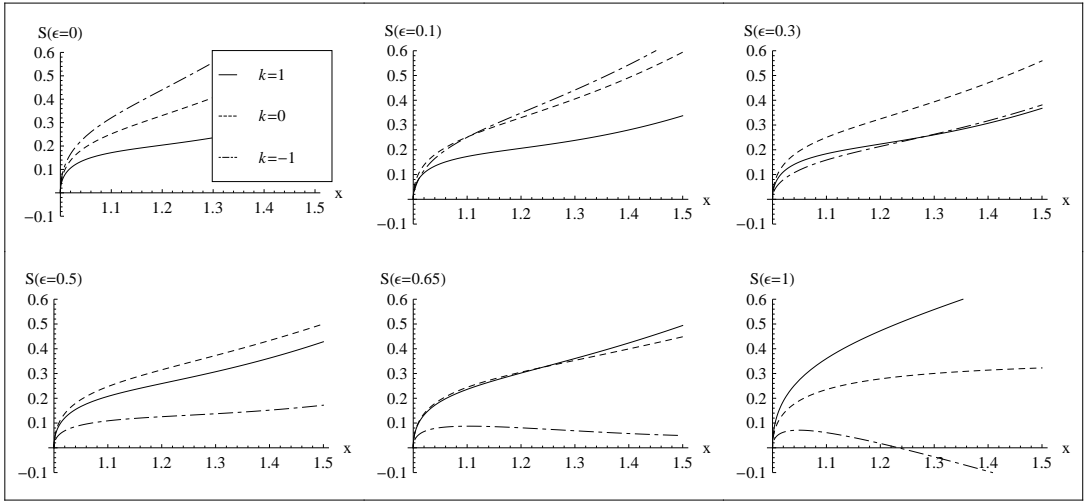}
\caption{Seiberg-Witten actions for uncharged $k=+1$, $k=0$, and $k=-1$ black holes are represented as solid curve, dotted curve and dash-dotted curve, respectively. The values of $\epsilon$ from top left to top right are 0, 0.1, and 0.3; while the values from bottom left to bottom right are 0.5, 0.65, and 1 (which reduces to general relativity). The horizons are set to unity. }
\end{center}
\end{figure}

\begin{figure}
\begin{center}
\includegraphics[width=4.8 in]{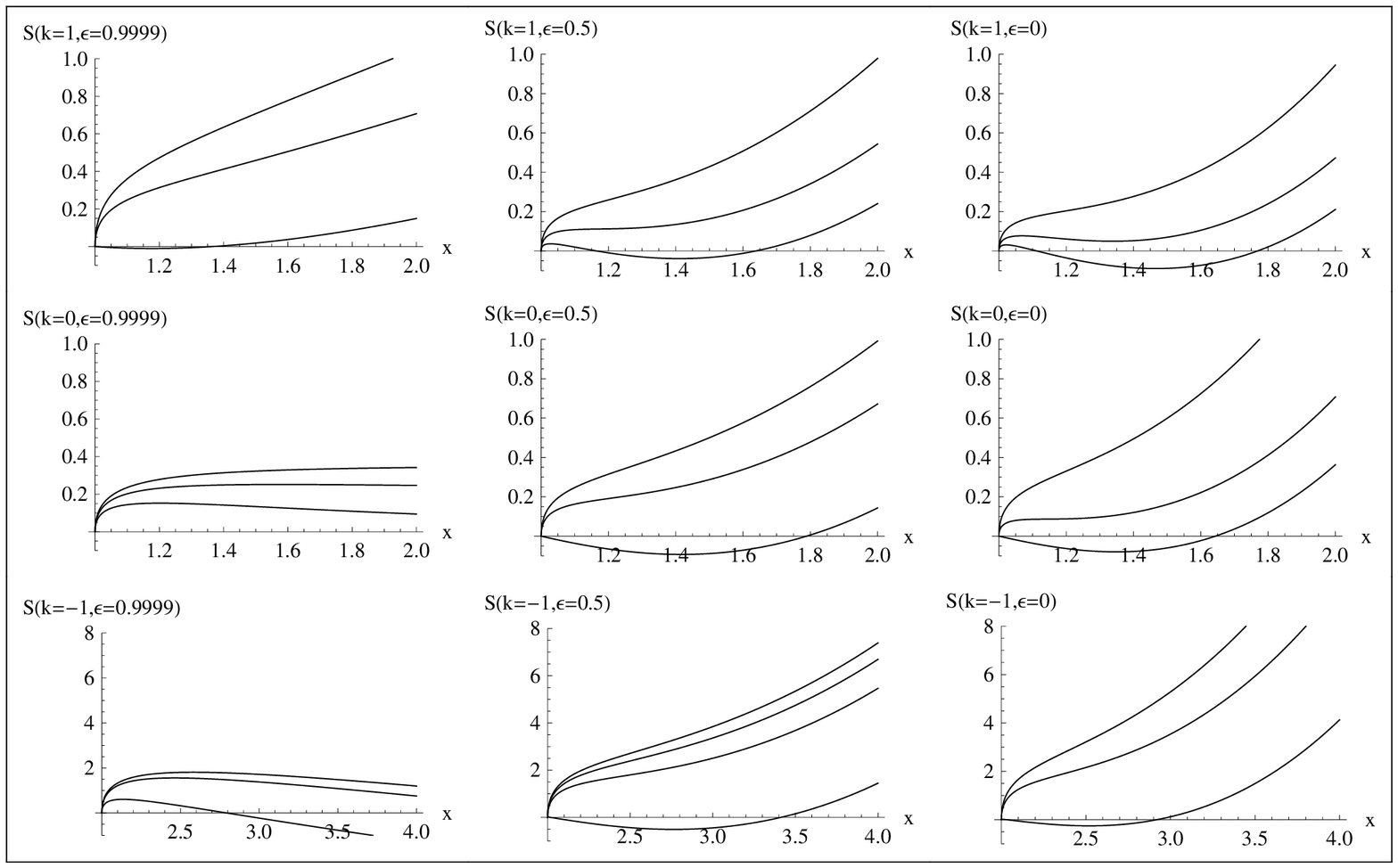}
\caption{The behavior of Seiberg-Witten actions as the charge increases. The first, second and third row correspond to $k=+1$, $k=0$ and $k=-1$ cases, respectively; while the first, second and third column correspond to $\epsilon=0.9999$, $\epsilon=0.5$ and $\epsilon=0$, respectively. Higher curves correspond to lower electrical charge. Horizons are set to unity for $k=1$ and $k=0$ cases while those for $k=-1$ black holes are set to 2. The asymptotic behavior for $\epsilon \to 1$ is however \emph{misleading} in this plot: The graphs for $S(k=0,\epsilon=0.9999)$ and $S(k=-1,\epsilon=0.9999)$ actually diverges to infinity, as shown by the plot with bigger scale in Fig.5.}
\end{center}
\end{figure}

\section{Discussion}

We have studied the Seiberg-Witten stability issues of asymptotically AdS Ho\v{r}ava-Lifshitz black holes and found that for all three types of topological black holes $(k=-1,0,+1)$ there exists a parameter range in which the Seiberg-Witten action dips below the $x$ axis for intermediate values of $x$, where $x=r\sqrt{-\Lambda}$ and $r$ is the radial coordinate, but pulls back up and diverges to infinity as $x \to \infty$. This means that although some of these black holes are unstable in the Seiberg-Witten sense, there exist ``nearby solutions'' which are stable in the Maldacena-Maoz sense  \cite{MaldacenaMaoz}.  At this stage, the nature of these nearby solutions and their possible implications on the dual field theories are not clear, which we leave open for future works to explore. In particular, we have been arguing rather naively that such nearby solutions exist and this needs to be proved. After all, not every naive deformation of a black hole can be stable; without knowing the details it is hard to guarantee at this point that there exists any stable deformation in view of  \cite{McInnes5}. As stability is inversely proportional to how much of the action is negative, we conjecture that:

\begin{itemize}
\item[(1.)] Uncharged positively curved AdS Ho\v{r}ava-Lifshitz black holes are stable in the Seiberg-Witten sense, while charged ones can be stable in Seiberg-Witten sense if the action remains positive, or stable in the Maldacena-Maoz sense if the action becomes negative for some finite region. For sufficiently charged black holes, stability is less guaranteed as $\epsilon \to 0$.
\item[(2.)] Uncharged flat AdS Ho\v{r}ava-Lifshitz black holes are stable in the Seiberg-Witten sense. Charged ones can be stable in Maldacena-Maoz sense if part of the Seiberg-Witten action becomes negative, but this stability is less guaranteed as $\epsilon \to 1$. 
\item[(3.)] Uncharged negatively curved AdS Ho\v{r}ava-Lifshitz black holes are stable in the Seiberg-Witten sense as long as $\epsilon$ is not too close to unity (For black holes with horizon set at $x$=2, the action starts to have a negative portion at $\epsilon \gtrsim 0.98$. The lower bound on $\epsilon$ before action starts to develop negative part becomes larger for larger black hole.) Charged ones can be stable in the Maldacena-Maoz sense if part of the Seiberg-Witten action becomes negative, but again this stability is less guaranteed as $\epsilon \to 1$. 
\end{itemize}

It is unlikely that in cases 2 and 3 the black hole spacetime can be stable for $\epsilon$ close to 1, even in the Maldacena-Maoz sense that there exists a nearby solution; instead there might exist a critical value of $\epsilon$ for a given value of charge such that the black hole spacetime becomes genuinely unstable even in the Maldacena-Maoz sense. One would need to know the exact ways the black hole spacetime changes in response to brane nucleations to know if this is indeed the case, and if so to find such critical value. Lacking a quantitative way to investigate this issue right now, we leave this problem for future investigation.

At this point, we would like to remind the readers that our analysis should be taken with skepticism, in view of the caveats we mentioned in the introduction. Nevertheless, if our analysis is at least qualitatively correct, it provides an example of instability with the Seiberg-Witten action being \emph{eventually positive} but which is not bounded below as we approach $\epsilon=1$. This recovers a general relativistic result of Seiberg-Witten instability, which, though not a proof of validity of our approach, is at least a consistency check.  

We now note that our analysis is somewhat incomplete since in the application to holographic superconductors, it is typical to add not just a Maxwell field  \cite{cai-zhang}  \cite{jwc}, but also a scalar field. Therefore it is possible that in the cases where the Seiberg-Witten action becomes negative for some finite range of $x$, the solution can nevertheless be stabilized when the scalar field is taken into account, even without deforming the black hole. Note that in  \cite{cai-zhang} and  \cite{jwc}, the black holes considered are \emph{uncharged}, with the background Maxwell field and scalar field weakly coupled to gravity, i.e. there is no backreaction to the metric. There is no Seiberg-Witten instability for all values of $\epsilon$ for flat uncharged black holes so all are fine in such applications, which is what motivated our investigation in the first place.

\begin{figure}
\begin{center}
\includegraphics[width=5.0 in]{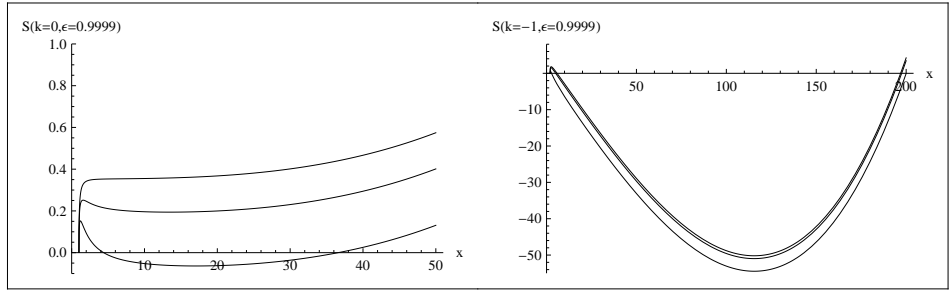}
\caption{The Seiberg-Witten action for $S(k=0,\epsilon=0.9999)$ and $S(k=-1,\epsilon=0.9999)$ diverge to infinity.}
\end{center}
\end{figure}

\begin{figure}
\begin{center}
\includegraphics[width=5.0 in]{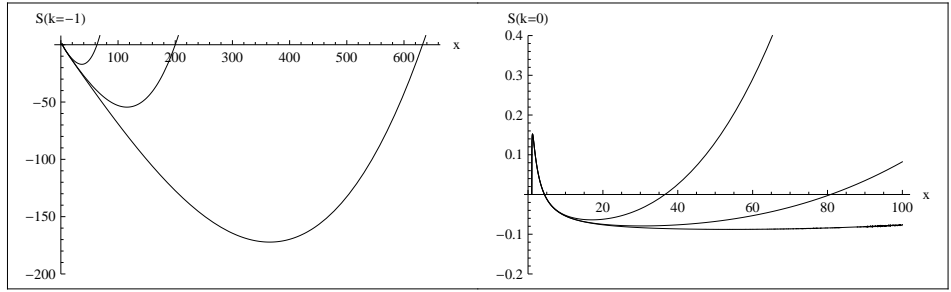}
\caption{The actions $S(k=0)$ and $S(k=-1)$ are not bounded below as $\epsilon \to 1$. The curves from top to bottom for $S(k=-1) $correspond to $\epsilon=0.999$, $\epsilon=0.9999$ and $\epsilon=0.99999$ respectively; while the curves from top to bottom for $S(k=0)$ correspond to $\epsilon=0.99999$,$\epsilon=0.999999$ and $\epsilon=0.9999999$ respectively. The actions eventually recover their corresponding behaviors in general relativity as $\epsilon=1$ and turning point is pushed to infinity.}
\end{center}
\end{figure}

Nevertheless, recall that in analysis of holographic superconductors dual to Einstein-Maxwell-scalar black holes, we expect the black holes to develop scalar hair around and below a certain critical temperature $T_c$ if the Breitenlohner-Freedman bound is violated near the horizon of near-extremal black holes  \cite{Hartnoll}  \cite{Horowitz}  \cite{HorowitzRoberts}.  Therefore it will be interesting to explore the stability issue when both the Maxwell field and scalar field are strongly coupled to gravity in Ho\v{r}ava-Lifshitz theory. We expect that these black holes would exhibit behavior similar to dilaton black holes and as such stable in the Seiberg-Witten sense  \cite{yenchin}, although how much this behavior is preserved under Ho\v{r}ava-Lifshitz gravity is yet to be explored.


Finally we recall that the present work only considers Ho\v{r}ava-Lifshitz black holes for $\lambda=1$. The case for general $\lambda$ is not only difficult from a calculation point of view, but also conceptually: the metrics are no longer asymptotically anti--de Sitter  \cite{CCO}, and it is not clear whether ``AdS''/CFT makes sense for such black holes (though perhaps not entirely impossible -- see e.g. so-called non-AdS/non-CFT correspondence  \cite{Ofer}). It might also be interesting to explore the Seiberg-Witten instability for Ho\v{r}ava-Lifshitz black holes in higher dimensions (for a study of five-dimensional Ho\v{r}ava-Lifshitz black holes, see  \cite{KPPT}). 

It is clear that Ho\v{r}ava-Lifshitz black holes, especially in the case without detailed balance have rich physics in need of further study. We may be able to learn something interesting even if Ho\v{r}ava-Lifshitz gravity turns out not to be a theory for describing gravity in our world. 

\acknowledgments
We thank Brett McInnes from National University of Singapore for fruitful discussions on Seiberg-Witten instability. Pisin Chen is supported by Taiwan National Science Council under Project No. NSC 97-2112-M-002-026-MY3, by Taiwan's National Center for Theoretical Sciences, and by US Department of Energy under Contract No. DE-AC03-76SF00515. Yen Chin Ong is supported by Taiwan Scholarship from Taiwan's Ministry of Education.


\begin{thebibliography}{100}

\bibitem{Maldacena} J. M. Maldacena, \emph{The Large N Limit of Superconformal Field Theory and Supergravity}, Adv. Theor. Math. Phys.2:231-252, 1998, \href{http://arxiv.org/abs/hep-th/9711200}{[arXiv:hep-th/9711200v3]}.

\bibitem{Gubser1} S. S. Gubser, I. R. Klebanov and A. M. Polyakov, \emph{Gauge Theory Correlators from Non-Critical String Theory}, Phys. Lett. B 428, 105 (1998), \href{http://arxiv.org/abs/hep-th/9802109}{[arXiv:hep-th/9802109v2]}.

\bibitem{Witten1} E. Witten, \emph{Anti de Sitter Space and Holography}, Adv.Theor.Math.Phys.2:253-291,1998, \href{http://arxiv.org/abs/hep-th/9802150}{[arXiv:hep-th/9802150v2]}.

\bibitem{Chamblin} A. Chamblin, R. Emparan, C. V. Johnson, Robert C. Myers, \emph{Holography, Thermodynamics and Fluctuations of Charged AdS Black Holes}, Phys.Rev.D60:104026, 1999, \href{http://arxiv.org/abs/hep-th/9904197}{[arXiv:hep-th/9904197v1]}.

\bibitem{McInnes1}  B. McInnes, \emph{Bouding the Temperatures of Black Holes Dual to Strongly Coupled Field Theories on Flat Spacetime}, JHEP09(2009)048, \href{http://arxiv.org/abs/0905.1180}{[arXiv:0905.1180v3 [hep-th]]}.

\bibitem{McInnes2} B. McInnes, \emph{Holography of the Quark Matter Triple Point}, Nuclear Physics B832 (2010) 323-341, \href{http://arxiv.org/abs/0910.4456}{[arXiv:0910.4456v3 [hep-th]]}.

\bibitem{Hartnoll} S. A. Hartnoll, C. P. Herzog and G. T. Horowitz, \emph{Holographic Superconductors}, JHEP 0812,015 (2008), \href{http://arxiv.org/abs/0810.1563}{[arXiv:0810.1563v1 [hep-th]]}.

\bibitem{Debu} C-Y. Huang, F-L. Lin, D. Maity, \emph{Holographic Multi-Band Superconductor}, Phys.Lett.B703:633-640,2011, \href{http://arxiv.org/abs/1102.0977}{[arXiv:1102.0977v2 [hep-th]]}

\bibitem{Christopher} C. P. Herzog, \emph{Lectures on Holographic Superfluidity and Superconductivity}, J. Phys. A42: 343001, 2009, \href{http://arxiv.org/abs/0904.1975}{[arXiv:0904.1975v2 [hep-th]]}.

\bibitem{Horowitz} G. Horowitz, \emph{Introduction to Holographic Superconductors},  ``From Gravity to Thermal Gauge Theories: the AdS/CFT Correspondence'', p. 313, Springer, 2011, \href{http://arxiv.org/abs/1002.1722}{[arXiv:1002.1722v2 [hep-th]]}.

\bibitem{Subir1} S. Sachdev, \emph{Holographic metals and the fractionalized Fermi liquid}, Physical Review Letters 105, 151602 (2010), \href{http://arxiv.org/abs/1006.3794}{[arXiv:1006.3794v3 [hep-th]]}.

\bibitem{Subir2} S.Sachdev, \emph{Strange metals and the AdS/CFT correspondence}, J. Stat. Mech. 1011:P11022, 2010, \href{http://arxiv.org/abs/1010.0682}{[arXiv:1010.0682v3 [cond-mat.str-el]]}.

\bibitem{Puletti} V. Giangreco M. Puletti, S. Nowling, L. Thorlacius, T. Zingg, \emph{Holographic Metals at Finite Temperature}, JHEP 1101:117 (2011), \href{http://arxiv.org/abs/1011.6261}{[arXiv:1011.6261v2 [hep-th]]}. 

\bibitem{Hartnoll2} S. A. Hartnoll, \emph{Lectures on Holographic Methods for Condensed Matter Physics}, Class.Quant.Grav.26:224002, 2009, \href{http://arxiv.org/abs/0903.3246}{[arXiv:0903.3246v3 [hep-th]]}. 

\bibitem{SeibergWitten} N. Seiberg and E. Witten, \emph{The D1/D5 System and Singular CFT}, JHEP 9904(1999)017, \href{http://arxiv.org/abs/hep-th/9903224}{[arXiv:hep-th/9903224v3]}.

\bibitem{cai-zhang} R-G. Cai, H-Q. Zhang, \emph{Holographic Superconductors with Ho\v{r}ava-Lifshitz Black Holes}, Phys.Rev.D81:066003, 2010, \href{http://arxiv.org/abs/0911.4867}{[arXiv:0911.4867v2 [hep-th]]}.

\bibitem{jwc} J. Jing, L. Wang, S. Chen, \emph{Holographic Superconductors in $z=3$ Ho\v{r}ava-Lifshitz gravity without condition of detailed balance}, \href{http://arxiv.org/abs/1001.1472}{[arXiv:1001.1472v4 [hep-th]]}.

\bibitem{Setare} M. R. Setare, N. Majd, D. Momeni, \emph{Holographic Superconductors in a Model of Non-Relativistic Gravity}, JHEP05(2011)118, \href{http://arxiv.org/abs/1003.0376}{[arXiv:1003.0376v3 [hep-th]]}. 

\bibitem{Zingg} E.J. Brynjolfsson, U.H. Danielsson, L. Thorlacius and T. Zingg, \emph{Holographic Superconductors with Lifshitz Scaling}, J.Phys.A43:065401,2010, \href{http://arxiv.org/abs/0908.2611}{[arXiv:0908.2611v3 [hep-th]]}.

\bibitem{Sin} S-J. Sin, S-S. Xu and Y. Zhou, \emph{Holographic Superconductor for a Lifshitz Fixed Point}, 	Int.J.Mod.Phys.A26:4617-4631,2011, \href{http://arxiv.org/abs/0909.4857}{[arXiv:0909.4857v4 [hep-th]]}.

\bibitem{Cremonesi} S. Cremonesi, D. Melnikov and Y. Oz, \emph{Stability of Asymptotically Schroedinger RN Black Hole and Superconductivity}, JHEP04 (2010) 048, \href{http://arxiv.org/abs/0911.3806}{[arXiv:0911.3806v1 [hep-th]]}.

\bibitem{Tatsuma} T. Nishioka, \emph{Ho\v{r}ava-Lifshitz Holography}, Class. Quant. Grav.26:242001,2009, \href{http://arxiv.org/abs/0905.0473v1}{[arXiv:0905.0473v1 [hep-th]]}.

\bibitem{Kluson} J. Kluso\v{n}, \emph{String in Ho\v{r}ava-Lifshitz Gravity}, Phys. Rev. D 82, 086007 (2010), \href{[http://arxiv.org/abs/1002.2849v3}{[arXiv:1002.2849v3 [hep-th]]}.

\bibitem{Kluson2} J. Kluso\v{n}, K. L. Panigrahi, \emph{T-Duality For String in Ho\v{r}ava-Lifshitz Gravity}, Eur. Phys. J. C 71, 1595 (2011), \href{http://arxiv.org/abs/1006.4530}{[arXiv:1006.4530v2 [hep-th]]}.

\bibitem{Strominger} I. Bredberg, C. Keeler, V. Lysov, A. Strominger, \emph{Cargese Lectures on the Kerr/CFT Correspondence}, Nucl. Phys. B, Proc. Suppl. 216, 194 (2011), \href{http://arxiv.org/abs/1103.2355}{[[arXiv:1103.2355v3 [hep-th]]}.

\bibitem{horava0} P. Horava, \emph{Membranes at Quantum Criticality}, JHEP 0903:020, 2009, \href{http://arxiv.org/abs/0812.4287}{[arXiv:0812.4287v3 [hep-th]]}.

\bibitem{Kluson3} J. Kluso\v{n}, \emph{Branes at Quantum Criticality}, JHEP 0907:079, 2009, \href{http://arxiv.org/abs/0904.1343v4}{[arXiv:0904.1343v4 [hep-th]]}.


\bibitem{Nabamita} N. Banerjee, S. Dutta, D. P. Jatkar, \emph{Geometry and Phase Structure of Non-Relativistic Branes}, Class.Quant.Grav.28:165002,2011, \href{http://arxiv.org/abs/1102.0298}{[arXiv:1102.0298v3 [hep-th]]}.

\bibitem{Sekhar} S. Sekhar Pal, \emph{Non-Relativistic Supersymmetric Dp Branes}, Class. Quant. Grav. 26:245014,2009, \href{http://arxiv.org/abs/0904.3620}{[arXiv:0904.3620v4 [hep-th]]}.

\bibitem{Strominger2} I. Bredberg, C. Keeler, V. Lysov, A. Strominger, \emph{Wilsonian Approach to Fluid/Gravity Duality}, JHEP 1103:141, 2011, \href{http://arxiv.org/abs/1006.1902}{[arXiv:1006.1902v1 [hep-th]]}.

\bibitem{Laurent} L. Freidel, \emph{Reconstructing AdS/CFT}, \href{http://arxiv.org/abs/0804.0632v1}{[arXiv:0804.0632v1 [hep-th]]}.

\bibitem{Birmingham} D. Birmingham, \emph{Topological Black Holes in anti--de Sitter Space}, Class.Quant.Grav. 16 (1999) 1197-1205, \href{http://arxiv.org/abs/hep-th/9808032}{ [arXiv:hep-th/9808032v3]}.

\bibitem{Stefan} Ste. Aminneborg, I. Bengtsson, S. Holst, P. Peldan, \emph{Making anti--de Sitter Black Holes}, Class. Quant. Grav. 13 (1996) 2707-2714, \href{http://arxiv.org/abs/gr-qc/9604005}{[arXiv:gr-qc/9604005v1]}.

\bibitem{Kleban} M. Kleban, M. Porrati, R. Rabadan, \emph{Stability in Asymptotically AdS Spaces}, 	JHEP0508 (2005) 016, \href{http://arxiv.org/abs/hep-th/0409242}{[arXiv:hep-th/0409242v1]}.


\bibitem{Besse} A. L. Besse, \emph{Einstein Manifolds}, Classics in Mathematics, Springer 2007.

\bibitem{BF} P. Breitenlohner and D. Z. Freedman, \emph{Positive Energy in anti--de Sitter Backgrounds and Gauged Extended Supergravity}, Phys. Lett. B115 (1982) 197.

\bibitem{Lee} J. M. Lee, \emph{The Spectrum of an Asymptotically Hyperbolic Einstein Manifold}, Communications in Analysis and Geometry 3 (1995) 253-271, \href{http://arxiv.org/abs/dg-ga/9409003}{[arXiv:dg-ga/9409003v1]}.


\bibitem{Schwinger}  J. Schwinger, \emph{On Gauge Invariance and Vacuum Polarization}, Phys. Rev. 82: 664-679. 

\bibitem{Barbon} J.L.F. Barb\'on, J. Mart\'inez-Mag\'an, \emph{Spontaneous Fragmentation of Topological Black Holes}, JHEP08(2010)031, \href{http://arxiv.org/abs/1005.4439}{[arXiv:1005.4439v1 [hep-th]]}.

\bibitem{Schoen} R. Schoen, \emph{Conformal Deformation of a Riemannian Metric to Constant Scalar Curvature}, J. Diff. Geom. 20 (1984) 1.

\bibitem{Xue} W. Xue, \emph{Non-Relativistic Supersymmetry}, \href{http://arxiv.org/abs/1008.5102v1}{[arXiv:1008.5102v1 [hep-th]]}.

\bibitem{brett0} B. McInnes, \emph{Topologically Induced Instability in String Theory}, JHEP 0103:031, 2001, \href{http://arxiv.org/abs/hep-th/0101136}{[arXiv:hep-th/0101136v2]}.

\bibitem{Brett00} B. McInnes, \emph{Stringy Instability of Topologically Non-Trivial Ads Black Holes and of de Sitter S-Brane Spacetimes}, Nucl.Phys.B660:373-388, 2003, \href{http://arxiv.org/abs/hep-th/0205103}{[arXiv:hep-th/0205103v5]}.

\bibitem{Roman} R. A. Konoplya, A. Zhidenko, \emph{Stability of Higher Dimensional Reissner-Nordstr\"om-anti--de-Sitter Black Holes}, Phys. Rev.D78: 104017, 2008, \href{http://arxiv.org/abs/0809.2048}{[arXiv:0809.2048v1 [hep-th]]}.

\bibitem{GubserMitra} S. S. Gubser, Indrajit Mitra, \emph{Instability of Charged Black Holes in anti--de Sitter Space}, \href{http://arxiv.org/abs/hep-th/0009126}{[arXiv:hep-th/0009126v1]}.

\bibitem{McInnes3} B. McInnes, \emph{Black Hole Final State Conspiracies}, Nucl. Phys B807:33-55, 2009, \href{http://arxiv.org/abs/0806.3818}{[arXiv:0806.3818v2 [hep-th]]}.


\bibitem{GubserRocha} S. S. Gubser and F. D. Rocha, \emph{Peculiar Properties of a Charged Dilatonic Black Hole in $AdS_5$}, Phys. Rev. D 81:046001, (2010), \href{http://arxiv.org/abs/0911.2898}{[arXiv:0911.2898v2 [hep-th]]}.

\bibitem{Goldstein} K. Goldstein, S. Kachru, S. Prakash, and S. P. Trivedi, \emph{Holography of Charged Dilaton Black Holes}, JHEP 1008:078, 2010, \href{http://arxiv.org/abs/0911.3586}{[arXiv:0911.3586v4 [hep-th]]}.

\bibitem{Chiangmei} C-M. Chen and D-W. Pang, \emph{Holography of Charged Dilaton Black Holes in General Dimensions}, JHEP06 (2010) 093, \href{http://arxiv.org/abs/1003.5064}{[arXiv:1003.5064v2 [hep-th]]}.

\bibitem{Cadoni} M. Cadoni, P. Pani, \emph{Holography of Charged Dilatonic Black Branes at Finite Temperature}, JHEP 1104:049, 2011, \href{http://arxiv.org/abs/1102.3820v2}{[arXiv:1102.3820v2 [hep-th]]}.

\bibitem{Meyer} R. Meyer, B. Gouteraux, B. S. Kim, \emph{Strange Metallic Behaviour and the Thermodynamics of Charged Dilatonic Black Holes},  Fortschr. Phys. 59, 741 (2011), \href{http://arxiv.org/abs/1102.4433}{[arXiv:1102.4433v1 [hep-th]]}.

\bibitem{yenchin} Y. C. Ong, \emph{Stringy Stability of Dilaton Black Holes in 5--Dimensional anti--de Sitter Space}, Proceedings of the Conference in Honor of Murray Gell-Mann's 80th Birthday, p.583-590, World Scientific, 2010, Singapore, \href{http://arxiv.org/abs/1101.5776}{[arXiv:1101.5776v1 [hep-th]]}.


\bibitem{Horava1} P. Ho\v{r}ava, \emph{Quantum Gravity at a Lifshitz Point}, Phys.Rev.D79:084008, 2009, \href{http://arxiv.org/abs/0901.3775}{[arXiv:0901.3775v2 [hep-th]]}.

\bibitem{Thomas} T. P. Sotiriou, \emph{Ho\v{r}ava-Lifshitz Gravity: A Status Report}, J.Phys.Conf.Ser.283:012034, 2011, \href{http://arxiv.org/abs/1010.3218v2}{[arXiv:1010.3218v2 [hep-th]]}.

\bibitem{ADM} R. Arnowitt, S. Deser, C. W. Misner, \emph{The Dynamics of General Relativity}, ``Gravitation: an introduction to current research'', Louis Witten ed. (Wiley 1962), chapter 7, pp 227--265, \href{http://arxiv.org/abs/gr-qc/0405109}{[arXiv:gr-qc/0405109v1]}.

\bibitem{Chien-I} C-I. Chiang, J-A. Gu, P. Chen, \emph{Constraining the Detailed Balance Condition in Horava Gravity with Cosmic Accelerating Expansion}, JCAP 1010:015, 2010, \href{http://arxiv.org/abs/1007.0543}{[arXiv:1007.0543v2 [astro-ph]]}.

\bibitem{Horatiu} H. Nastase, \emph{On IR Solutions in Ho\v{r}ava Gravity Theories}, \href{http://arxiv.org/abs/0904.3604}{[arXiv:0904.3604v2 [hep-th]]}.

\bibitem{LMP} H. Lu, Jianwei Mei, C.N. Pope, \emph{Solutions to Horava Gravity}, Phys.Rev.Lett.103:091301, 2009, \href{http://arxiv.org/abs/0904.1595}{[arXiv:0904.1595v4 [hep-th]]}.

\bibitem{CCO} R-G. Cai, L-M. Cao, N. Ohta, \emph{Topological Black Holes in Ho\v{r}ava-Lifshitz Gravity}, Phys. Rev. D 80, 024003 (2009), \href{http://arxiv.org/abs/0904.3670}{[arXiv:0904.3670v3 [hep-th]]}.

\bibitem{Ehsan} A. Ghodsi, E. Hatefi, \emph{Extremal Rotating Solutions in Horava Gravity}, Phys Rev D.81.044016, \href{http://arxiv.org/abs/0906.1237}{[arXiv:0906.1237v3 [hep-th]]}.

\bibitem{Konoplya} R. A. Konoplya, \emph{Towards Constraining of the Horava-Lifshitz Gravities}, Phys.Lett.B679:499-503, 2009, \href{http://arxiv.org/abs/0905.1523}{[arXiv:0905.1523v5 [hep-th]]}.

\bibitem{Myung-Kim} Y. S. Myung, Y-W. Kim, \emph{Thermodynamics of Ho\v{r}ava-Lifshitz Black Holes}, Eur. Phys. J. C 68, 265 (2010), \href{http://arxiv.org/abs/0905.0179}{[arXiv:0905.0179v3 [hep-th]]}.

\bibitem{Greenwald} J. Greenwald, J. Lenells, J. X. Lu, V. H. Satheeshkumar, A. Wang, \emph{Black Holes and Global Structures of Spherical Spacetimes in Horava-Lifshitz theory}, 	Phys. Rev. D84, 084040 (2011) \href{http://arxiv.org/abs/1105.4259v3}{[arXiv:1105.4259v3 [hep-th]]}.

\bibitem{Kiritsis} E. Kiritsis, G. Kofinas, \emph{On Horava-Lifshitz ``Black Holes''}, JHEP01 (2010) 122\href{http://arxiv.org/abs/0910.5487v1}{[arXiv:0910.5487v1 [hep-th]]}. 

\bibitem{Padmanabhan} T. Padmanabhan, \emph{Lessons from Classical Gravity about the Quantum Structure of Spacetime}, J.Phys.Conf.Ser.306:012001,2011, \href{http://arxiv.org/abs/1012.4476}{[arXiv:1012.4476v2 [gr-qc]]}.

\bibitem{CCS} Q-J. Cao, Y-X. Chen, K-N. Shao, \emph{Black Hole Phase Transitions in Ho\v{r}ava-Lifshitz Gravity}, Phys.Rev.D83:064015, 2011, \href{http://arxiv.org/abs/1010.5044}{[arXiv:1010.5044v2 [hep-th]]}. 

\bibitem{Hawking} S. W. Hawking, Don N. Page, \emph{Thermodynamics of Black Holes in anti--de Sitter Space}, Commun. Math. Phys. 87, 577-588 (1983). 

\bibitem{CCO2} R-G. Cai, L-M. Cao, N. Ohta, \emph{Thermodynamics of Black Holes in Ho\v{r}ava-Lifshitz Gravity}, Phys.Lett.B679:504-509, 2009, \href{http://arxiv.org/abs/0905.0751}{[arXiv:0905.0751v3 [hep-th]]}.

\bibitem{HM} G. T. Horowitz, R. C. Myers, \emph{The AdS/CFT Correspondence and a New Positive Energy Conjecture for General Relativity}, Phys.Rev.D59:026005, 1998, \href{http://arxiv.org/abs/hep-th/9808079}{[arXiv:hep-th/9808079v1]}.

\bibitem{Surya} S. Surya, K. Schleich, D. M. Witt, \emph{Phase Transitions for Flat AdS Black Holes}, Phys.Rev.Lett. 86 (2001) 5231-5234, \href{http://arxiv.org/abs/hep-th/0101134}{[arXiv:hep-th/0101134v2]}.

\bibitem{limiao} M. Li, Y. Pang, \emph{Trouble with Ho\v{r}ava-Lifshitz Gravity}, JHEP 0908:015, 2009, \href{http://arxiv.org/abs/0905.2751}{[arXiv:0905.2751v4 [hep-th]]}.


\bibitem{McInnes4} B. McInnes, \emph{A Universal Lower Bound on the Specific Temperature of AdS-Reissner-Nordstr\"om Black Holes with Flat Event Horizons}, Nuclear Physics B, 848(2011), 474-489, \href{http://arxiv.org/abs/1012.4056}{[arXiv:1012.4056v2 [hep-th]]}.

\bibitem{MaldacenaMaoz} J. Maldacena, L. Maoz, \emph{Wormholes in AdS}, JHEP 0402 (2004) 053, \href{http://arxiv.org/abs/hep-th/0401024}{[arXiv:hep-th/0401024v2]}.

\bibitem{KehagiasSfetsos} A. Kehagias, K. Sfetsos, \emph{The Black Hole and FRW Geometries of Non-Relativistic Gravity}, Phys.Lett.B678:123-126, 2009, \href{http://arxiv.org/abs/0905.0477}{[arXiv:0905.0477v1 [hep-th]]}.

\bibitem{McInnes5} B. McInnes, \emph{Fragile Black Holes}, Nucl.Phys.B842:86-106, 2011, \href{http://arxiv.org/abs/1008.0231}{[arXiv:1008.0231v2 [hep-th]]}.

\bibitem{HorowitzRoberts} G.T. Horowitz, M. M. Roberts, \emph{Zero Temperature Limit of Holographic Superconductors}, JHEP 0911:015, 2009, \href{http://arxiv.org/abs/0908.3677}{[arXiv:0908.3677v1 [hep-th]]}.

\bibitem{Ofer} O. Aharony, \emph{The non-AdS/non-CFT Correspondence, or Three Different Paths to QCD}, \href{http://arxiv.org/abs/hep-th/0212193}{[arXiv:hep-th/0212193v3]}.

\bibitem{KPPT} G. Koutsoumbas, E. Papantonopoulos, P. Pasipoularides, M. Tsoukalas, \emph{Black Hole Solutions in 5D Horava-Lifshitz Gravity}, \href{http://arxiv.org/abs/1004.2289}{[arXiv:1004.2289v2 [hep-th]]}



\end{thebibliography}
\end{document}